\documentclass[twocolumn,superscriptaddress,amsmath,amssymb,aps,pra]{revtex4-2}

\usepackage{graphicx}
\usepackage{amssymb}
\usepackage{amsmath}
\usepackage{bbm}
\usepackage{physics}
\usepackage[colorlinks,urlcolor=blue,citecolor=blue,linkcolor=blue]{hyperref}
\usepackage{mathtools}

\DeclareMathOperator{\Lagr}{\mathcal{L}}

\begin{document}

\title{Numerical evaluation and robustness of the quantum mean force Gibbs state}

\author{Yiu-Fung Chiu}
\affiliation{SUPA, School of Physics and Astronomy, University of St Andrews, St Andrews, KY16 9SS, United Kingdom}

\author{Aidan Strathearn}
\affiliation{School of Mathematics and Physics, The University of Queensland, St Lucia, Queensland 4072, Australia}

\author{Jonathan Keeling}
\affiliation{SUPA, School of Physics and Astronomy, University of St Andrews, St Andrews, KY16 9SS, United Kingdom}

\begin{abstract}
    We introduce a numerical method to determine the Hamiltonian of Mean Force (HMF) Gibbs state for a quantum system strongly coupled to a reservoir.  The method adapts the Time Evolving Matrix Product Operator (TEMPO) algorithm to imaginary time propagation.  By comparing the real-time and imaginary-time propagation for a generalized spin-boson model, we confirm that the HMF Gibbs state correctly predicts the steady state.  We show that the numerical dynamics match  the polaron master equation at strong coupling.  We illustrate the potential of the imaginary-time TEMPO approach by exploring reservoir-induced entanglement between qubits.
\end{abstract}

\maketitle

\section{Introduction}

The laws of thermodynamics imply that when a system is brought in contact with a reservoir, it will reach thermal equilibrium with that reservoir.  In simple cases,  where no extra conservation laws apply this will lead the system to adopt a Gibbs state~\cite{callen1998thermodynamics}.
Traditional thermodynamics considers this situation in the limit where the coupling between system and reservoir is weak, so that the Gibbs state of the system is determined purely by the energies of the system states.  However, more generally, the coupling between the system and reservoir will change the energies of states, as can be described by the potential of mean force~\cite{Hanggi1990}.   In an ergodic system, the time average of the dynamics should equal the ensemble average, and thus the dynamics of the system strongly coupled to a reservoir should sample from this potential of mean force.

Such ideas can be straightforwardly generalized to quantum systems by considering the density matrix and the construction of the Hamiltonian of Mean Force (HMF)~\cite{Jarzynski2004,Campisi2009,trushechkin2021open}.
In equilibrium, the system density matrix takes the form
$\rho^{}_{\text{HMF}} = \Tr_R[\rho^{(SR)}]$ where the system+reservoir density matrix is $\rho^{(SR)}=e^{-\beta (H_S+H_{SR})}/\Tr[e^{-\beta (H_S+H_{SR})}]$. Here $H_S$ is the system Hamiltonian, and $H_{SR}$ denotes both the reservoir Hamiltonian and its coupling to the system.  The Hamiltonian of mean force is defined by
$\rho^{}_{\text{HMF}} \equiv e^{-\beta H_{\text{HMF}}}/\Tr[e^{-\beta H_{\text{HMF}}}]$.  This concept is very useful when considering the thermodynamics of small quantum systems strongly coupled to their environment~\cite{kosloff2013quantum,vinjanampathy2016quantum,trushechkin2021open}.   Recent work~\cite{Cresser2021,trushechkin2021open} has provided analytic expressions for the form of the HMF Gibbs state in the limits of weak and strong system--reservoir coupling.

While thermodynamics is a useful tool to find the stationary properties of ergodic systems, it cannot  predict the dynamics of how that state is reached.  For that, a widely used approach is to determine equations of motion for the system density matrix, known as quantum master equations~\cite{Breuer2002}.   It can be shown that when the weak coupling master equation is derived correctly, it recovers the HMF Gibbs state at weak coupling~\cite{mori2008dynamics,fleming2011accuracy,thingna2012generalized,subacsi2012equilibrium}.  In the strong-coupling limit, while general arguments of ergodicity should make the HMF Gibbs state robust,  the literature is less clear.  In particular, recent predictions~\cite{Goyal2019,Orman2020} for the steady state of a strong-coupling master equation differ from the strong-coupling form of the HMF Gibbs state~\cite{Cresser2021}.  To help address this question, \citet{Cresser2021} have shown that at strong coupling, in cases where the system--reservoir coupling is proportional to a single system operator $X$,
the Hamiltonian of Mean Force takes the projected form
$H_{\text{HMF}} = \sum_n |n\rangle\langle n| H_S |n\rangle\langle n|$, where $|n\rangle$ are the eigenstates of $X$.  

In this paper we use the numerical ``Time Evolving Matrix Product Operator'' approach (TEMPO)~\cite{Strathearn2018,Strathearn2019,TimeEvolvingMPO} to compare the dynamics of a system at intermediate--strong coupling to the predictions of the HMF Gibbs state.  
We use TEMPO to perform evolution in both real time (for the dynamics), and a modification of TEMPO to imaginary-time propagation to provide numerical results for the HMF Gibbs state at intermediate couplings.
We consider a simple example of a two-level system, in the generic situation where the system Hamiltonian does not commute with the the system--reservoir coupling.
We find that (as expected) the dynamics reaches the HMF Gibbs state in all cases, and that the timescale to reach this state becomes exponentially long as the system--reservoir coupling increases.  Such a result is consistent with a strong-coupling approximation based on a polaron master equation~\cite{McCutcheon2010,trushechkin2021quantum}.   By comparing the numerical results  to the predictions of polaron theory, we note that the timescale for thermalisation matches well.  As noted elsewhere~\cite{trushechkin2021quantum,trushechkin2021open}, the steady state of the polaron theory is equivalent to the projected ensemble of Ref.~\cite{Cresser2021}, and so matches the HMF Gibbs state  in the limit as system--reservoir coupling goes to infinity.

As we explain further below, evaluating the HMF Gibbs state via imaginary-time TEMPO is significantly less computationally demanding than real-time evolution.  It thus provides an ideal method to find the properties of systems for intermediate system--reservoir coupling, where the analytical approximations at weak- and strong-coupling fail.  As an illustration of this, we analyze the steady state of two qubits, coupled to a common reservoir.  Such a model shows the intriguing behavior that qubit--qubit correlations vanish for both weak and strong coupling to the reservoir, but significant correlations exist for intermediate couplings.

The remainder of this article is organized as follows.  Section~\ref{sec:ImagTimeTEMPO} discusses the application of the TEMPO algorithm in imaginary time.   We then use this in Sec.~\ref{sec:OneQubit} to compare the dynamics and ensemble averaged populations for a single qubit.  Section~\ref{sec:TwoQubit} uses the imaginary time approach to study reservoir-induced coherence and entanglement between two qubits.  We then conclude in Sec.~\ref{sec:Conclusions}.  The Appendix presents details of the numerical convergence of the algorithms.

\section{Imaginary-time TEMPO}
\label{sec:ImagTimeTEMPO}

The TEMPO method is a reformulation of the iterative quasi-adiabatic propagator path integral (QUAPI) approach developed by \citet{Makri1995,Makri1995a}.  This numerical approach allows one to simulate a system with linear coupling to a reservoir of harmonic oscillators, with arbitrary forms and strengths of system--reservoir coupling.  The method has been applied to a variety of problems in open quantum systems~\cite{minoguchi2019environment,Gribben2020,Fux2021,BundgaardNielsen2021,popovic2021quantum,gribben2021using,bose2021tensor,bose2021multisite,richter2021enhanced}.
It has also been used to to make conceptual connections to the process tensor (PT) formulation of open quantum systems~\cite{Jorgensen2019,Lerose2021prx}.
More recently it has been understood as but one of a family of methods based on producing matrix product operator (MPO) representations of the process tensor~\cite{Jorgensen2019,Fux2021}. This PT-MPO formalism also allows one to devise alternate numerical methods to find the MPO representation of the PT for more general forms of reservoir~\cite{ye2021constructing,cygorek2021numerically}.  In the following we will however be focused on the original formalism of TEMPO as a time evolving algorithm.

For the real-time calculations presented below we make use of the open-source implementation of TEMPO~\cite{TimeEvolvingMPO}.  To determine the HMF Gibbs state, we instead use a new method, performing time evolution in imaginary time.  Specifically, we wish to evaluate the unnormalized reduced density matrix $\tilde{\rho}=\Tr_R\left[e^{-\beta (H_S+H_{SR})}\right]$ which we can then normalize at the end of the calculation. The operator $e^{-\beta (H_S+H_{SR})}$ be regarded as time evolution by a time $-i\beta$. Following the standard method~\cite{Makri1995,Makri1995a,Strathearn2018,Strathearn2019}, we divide this into $N$ imaginary timesteps of length $\Delta=\beta/N$ and use a Trotter splitting to separate the system and reservoir parts such that,
\begin{equation}\label{eq:trot}
    e^{-\beta (H_S+H_{SR})}= (e^{-\Delta H_S}e^{-\Delta H_{SR}})^N,
\end{equation}
which is correct to order $\Delta^2$. We assume the system couples linearly to a reservoir of bosonic modes such that
\begin{equation}\label{eq:coup}
   H_{SR}=\sum_i
     g_i X (b_i+b^\dagger_i) +\nu_i b^\dagger_i b_i,
\end{equation}
where $b_i$ are the bosonic reservoir modes, $\nu_i$ are their frequencies, and $X$ is a system operator. Expanding the system operator in terms of its eigenbasis, $X=\sum_j X_j \ket{j}\bra{j}$, at each timestep in Eq.~\eqref{eq:trot}, we can analytically trace out the Gaussian reservoir to obtain a path sum
\begin{equation}\label{eq:dpi}
  [\tilde{\rho}]_{j_0j_N}=Z_R\sum_{j_1,...j_{N-1}}\prod_{k=0}^{N-1}[e^{-\Delta \tau H_S}]_{j_{k}j_{k+1}}\prod_{k^\prime=0}^{k}I_{kk^\prime}(j_{k^\prime}, j_{k}),
\end{equation}
where $[O]_{jj'}=\bra{j}O\ket{j'}$ and $Z_R=\mathrm{Tr}[e^{\sum_i \nu_i b^\dagger_i b_i}]$ is the free reservoir partition function. Here we have defined the imaginary time influence functions
\begin{equation}
  I_{kk^\prime}(j_{k^\prime}, j_{k})=\exp(\eta_{kk^\prime}X_{j_{k}}X_{j_{k^\prime}}),  
\end{equation}
where the coefficients
\begin{equation}
\eta_{kk^\prime}=
\begin{dcases}
    \int_{k\Delta}^{(k+1)\Delta}d\tau \int_{k^\prime\Delta}^{(k^\prime+1)\Delta}d\tau'K(\tau-\tau') & k \ne k^\prime \\
    \int_{k\Delta}^{(k+1)\Delta}d\tau \int_{k^\prime\Delta}^{\tau}d\tau'K(\tau-\tau') & k = k^\prime 
\end{dcases}
\end{equation}
are defined in terms of the reservoir imaginary-time correlation function
\begin{equation}\label{eq:itcf}
    K(\tau)=\int_0^\infty d\nu J(\nu) \frac{\cosh(\beta \nu/2 - \tau\nu)}{\sinh(\beta \nu/2)}
\end{equation}
where $J(\nu)=\sum_i g_i^2 \delta(\nu-\nu_i)$ is the spectral density. 

\begin{figure}
    \centering
    \includegraphics[width=0.9\linewidth]{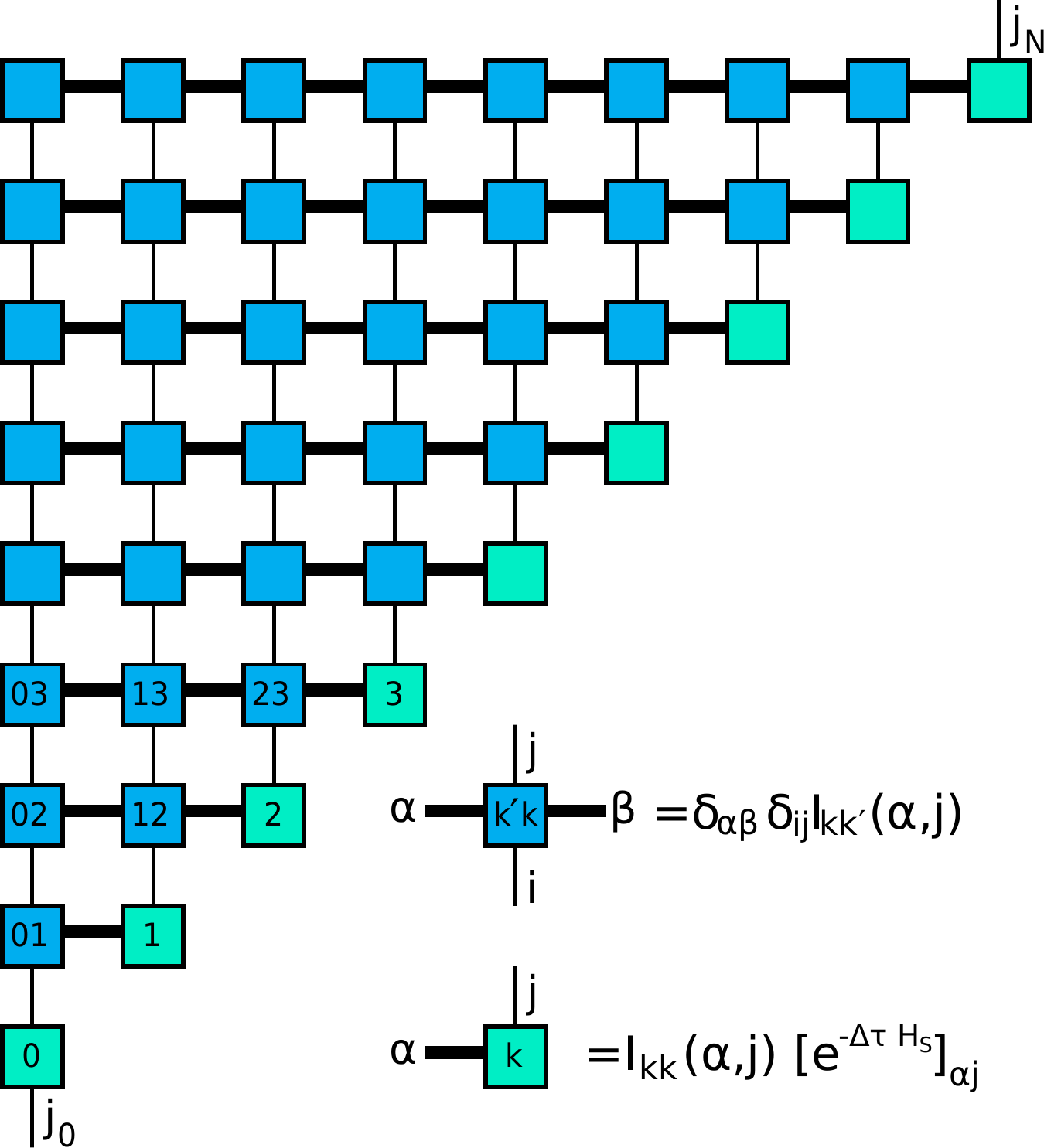}
    \caption{Tensor network representation of the summation written in Eq.~\eqref{eq:dpi}.  The network is contracted according to standard tensor-network contraction rules, in that lines connecting boxes indicate summation over that line.  The forms of each tensor are indicated in the bottom right.  In the bulk, the Kronecker deltas indicate that the tensors just pass the horizontal and vertical indices through---this is a consequence of working in the eigenbasis of $X$.    The symbols on the diagonal differ from those elsewhere in that they include propagation with the system Hamiltonian, which is not diagonal.}
    \label{fig:network}
\end{figure}

Similarly to real-time TEMPO, the discrete path sum in Eq.~\eqref{eq:dpi} can be written as a tensor network, depicted in Fig.~\ref{fig:network}, and contracted efficiently by decomposing it into a series of matrix product operators. For real-time propagation it was possible to make use of a finite memory cutoff to restrict the size of the network, since the real-time reservoir correlation function decays to zero. A finite memory cutoff is not possible for the imaginary-time correlation function, Eq.~\eqref{eq:itcf}, since this does not decay but instead satisfies, $K(\beta)=K(0)$. However, this potential increase in network size is offset by the fact that the size of the tensors here are $\mathcal{O}(d^4)$ for a $d$-dimensional system. In contrast, real-time propagation requires tensors of size $\mathcal{O}(d^8)$.  This difference occurs because in the imaginary-time approach $j=1\ldots d$, enumerating states in the Hilbert space, while for the real-time approach $j=1 \ldots d^2$, describing states in a doubled Hilbert space, corresponding to operators to the left and right of the density matrix.

In practice we ignore the factor of $Z_R$ in Eq.~\eqref{eq:dpi} when performing calculations so that the quantity we actually calculate is $\tilde{\rho}/Z_R$, which is still unnormalized. To normalize this and obtain a physical density matrix we divide by the trace.  The trace we obtain gives us the quantity $Z_{SR}/Z_{R}$, where $Z_{SR}$ is the full system-reservoir partition function. We note that by combining knowledge of the ratio $Z_{SR}/Z_{R}$ (which we obtain from our calculations) and of the free reservoir partition function, $Z_R$ (which is in principle analytically calculable) one could calculate thermodynamic quantities of the combined system-reservoir.

\section{Dynamics of the generalized qubit-boson model}
\label{sec:OneQubit}

In this section we compare the time evolution of a quantum system strongly coupled to an environment, as predicted by TEMPO~\cite{Strathearn2018,TimeEvolvingMPO}, against various thermodynamic ensembles.  We do this using a simple generalized spin-boson model, as also studied by~\citet{Cresser2021}. Our system Hamiltonian takes the form
$H_S=({\omega_q}/{2})\sigma^z$, in terms of system Pauli operators $\sigma^{x,y,z}$, while the reservoir Hamiltonian and coupling is as in Eq.~\eqref{eq:coup} with system coupling operator $X=\cos\theta \sigma^z-\sin\theta\sigma^x$.  
It is convenient to rotate our spin basis by $\theta$, so that we define
$\tau^z=\cos\theta \sigma^z-\sin\theta\sigma^x$,
$\tau^x=\cos\theta\sigma^x+\sin\theta \sigma^z$. This makes the pointer-state basis coincide with the $\tau^z$ eigenstates, and makes the model appear similar to standard spin-boson models: $H_S=(\omega_q/2)(\cos\theta\tau^z+\sin\theta\tau^x)$, and system-reservoir coupling $X=\tau^z$. As above, the reservoir is characterized by its spectral density $J(\nu)$; in what follows we parameterize $J(\nu)=\alpha \nu e^{-\nu/\omega_c}$.  We choose $\omega_q=k_BT=1, \omega_c=10$ in all our results, and vary $\alpha,\theta$.

\subsection{Robustness of the HMF Gibbs state}

We will compare the time evolution to the steady-state predictions of three ensembles: (1) The HMF Gibbs state $\rho_{\text{HMF}}$, found by imaginary-time TEMPO. (2) The system Gibbs state,  $\rho_{\text{sys}}=e^{-\beta H_S}/\Tr[e^{-\beta H_S}]$. (3) The projected Gibbs state 
$\rho_{\text{proj}}=e^{-\beta H_{\text{proj}}}/\Tr[e^{-\beta H_\text{proj}}]$, where $H_\text{proj}=\sum_\tau |\tau\rangle\langle \tau| H_S |\tau\rangle\langle \tau|$ where $|\tau\rangle$ is an eigenstate of $X=\tau^z$.  As discussed in the introduction, \citet{Cresser2021} have shown that at large coupling $\rho_{\text{HMF}}=\rho_{\text{proj}}$. For intermediate coupling however these  differ.

We focus in the following on the time evolution of the population in the pointer-state basis, $\langle\tau^z\rangle$ (technically the \emph{difference} in populations of the $\tau^z=\pm 1$ states). This is because the value of $\langle \tau^z \rangle$ distinguishes between different theories of the strong coupling steady state.  [In contrast, the pointer-state-basis coherence $\langle \tau^x \rangle$ is agreed to vanish in all strong-coupling theories, and this is indeed seen numerically (not shown).]
For the system and projected Gibbs states, analytic forms of $\langle \tau^z \rangle$ exist.  For $\rho_{\text{sys}}$, the system Hamiltonian, and thus the density matrix is diagonal in the original $\sigma^z$ basis, with elements $e^{\pm \beta \omega_q/2}/[2 \cosh(\beta \omega_q/2)]$. Rotating this into the $\tau^z$ basis gives
\begin{equation} \label{eq1}
\langle \tau^z\rangle_\text{sys} = - \cos(\theta) \tanh\left( \frac{\beta \omega_q}{2} \right).
\end{equation}
For $\rho_{\text{proj}}$ we have $H_{\text{proj}}=(\omega_q /2) \cos\theta \tau^z$ and thus:
\begin{equation}
\langle \tau^z\rangle_\text{proj} =
-\tanh\left(\frac{\beta \omega_{q}}{2} \cos \theta \right).\label{eq11}
\end{equation}

As noted in the introduction, there is disagreement in the literature for the steady state expected in the strong-coupling limit. The results of \citet{Orman2020} suggest a strong-coupling steady state density matrix of the form $\rho_{\text{OK}}=\sum_\tau |\tau\rangle\langle \tau| \rho_{\text{sys}} |\tau\rangle\langle \tau|$, with states $|\tau\rangle$ defined as above.  This projection has no effect on $\langle \tau^z \rangle$, so Ref.~\cite{Orman2020} would predict $\lim_{t\to\infty}\langle \tau^z(t)\rangle =\langle \tau^z\rangle_\text{sys}$. In contrast Refs.~\cite{Jarzynski2004,Campisi2009,Cresser2021} predict a steady state of $\rho_{\text{HMF}}$ giving $\lim_{t\to\infty}\langle \tau^z(t)\rangle=\langle \tau^z\rangle_\text{HMF}$, and Ref.~\cite{Cresser2021} showed that at strong coupling, $\langle \tau^z \rangle_{\text{HMF}}=\langle \tau^z\rangle_\text{proj}$.

\begin{figure}[ht]
    \includegraphics[width=\linewidth]{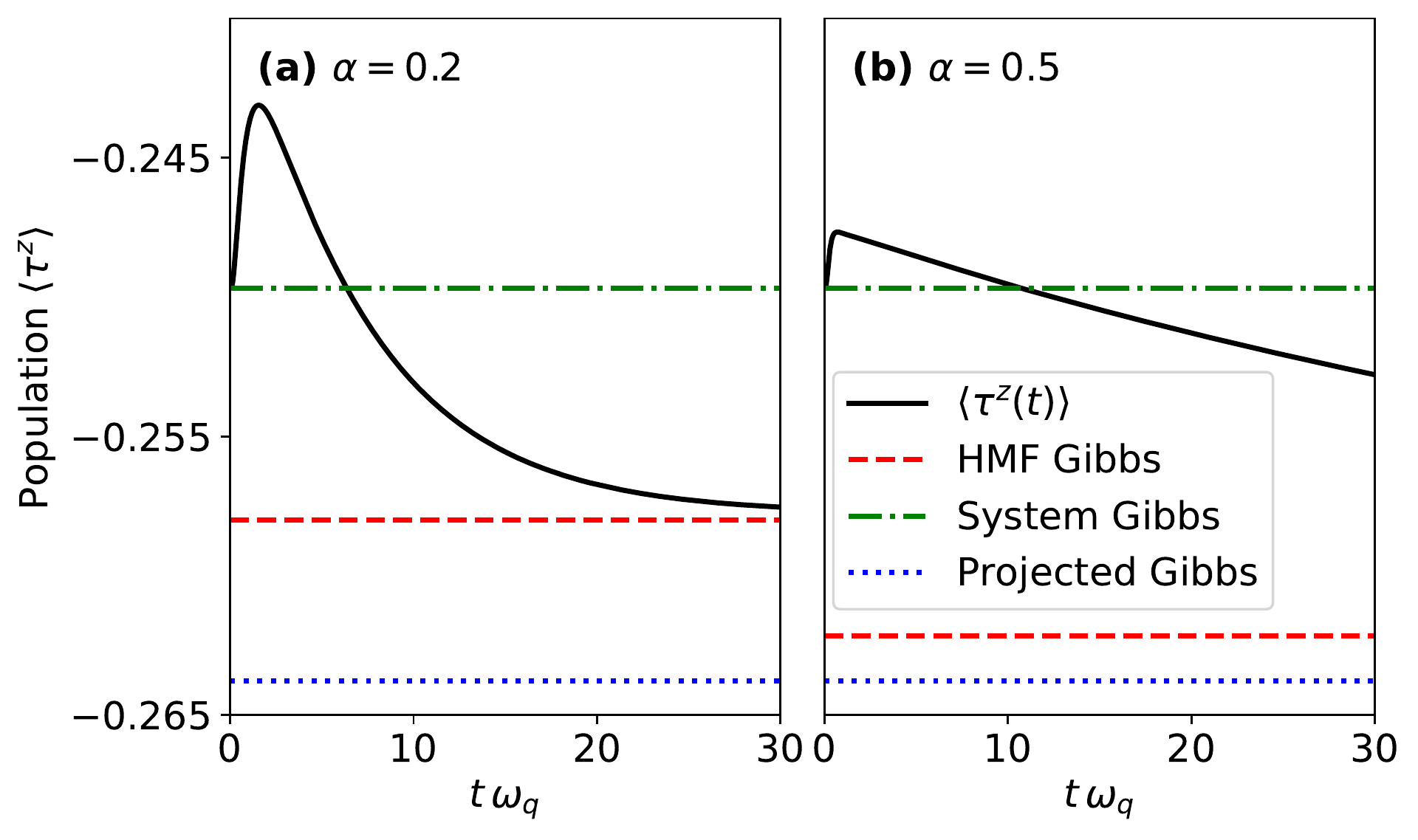}
    \caption{Time evolution of the population difference in the pointer basis, $\langle \tau^z \rangle$, after initializing in the system Gibbs state (black solid line) vs equilibrium predictions of various ensembles.  The ensembles shown are the HMF Gibbs state (red dashed), system Hamiltonian Gibbs state, Eq.~\eqref{eq1} (green dash dotted), and the Gibbs state for the system Hamiltonian projected onto the pointer basis, Eq.~\eqref{eq11} (blue dotted).  The latter two results do not depend on $\alpha$ and take the values $\langle \tau^z\rangle_{\text{sys}}=-0.250$ and $\langle \tau^z\rangle_{\text{proj}}=-0.264$ respectively.
    Results are shown for two values of the dimensionless system--reservoir coupling strength,  (a)  $\alpha = 0.2$ and (b) $\alpha = 0.5$.   Other parameters are $\theta=1$, $\omega_q=k_B T=1$ and $\omega_c=10$.}\label{Fig:1}
\end{figure}

Figure~\ref{Fig:1} shows the time evolution, after initial preparation in a factorized state with the system in the system Gibbs state. The coherence in pointer basis, $\langle \tau^x \rangle$ decays fast, so we do not show this.
Considering the population difference $\langle \tau^z \rangle$, we observe a fast initial change,  followed by a slow exponential decay to the HMF Gibbs state (as also discussed in the strong coupling limit by~\cite{trushechkin2021quantum}). We have checked (see further below) that the final state is the same no matter the choice of initial state.  Comparing the two panels, we see that, as discussed by~\citet{Cresser2021}, $\expval{\tau^z}_{\text{HMF}}$ approaches $\expval{\tau^z}_{\text{proj}}$ at strong coupling.  We also note that at stronger system--reservoir coupling, the late-time exponential decay becomes slower.   

\begin{figure}[ht]
    \includegraphics[width=\linewidth]{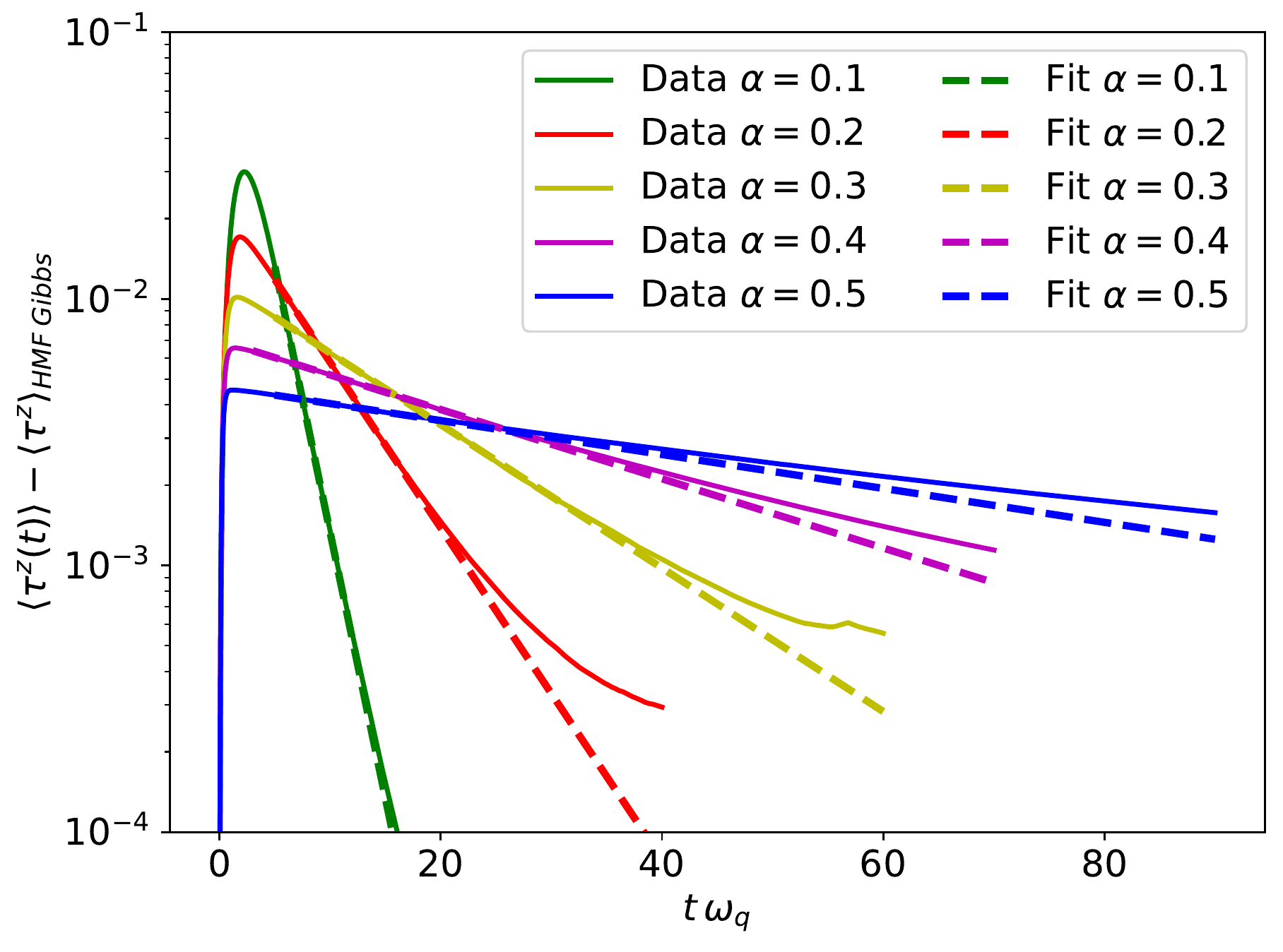}
    \caption{Difference between time evolution of population in pointer-state basis and the HMF Gibbs state as a log-linear plot.  The initial state here has the system prepared in the HMF Gibbs state, but neglects system--reservoir entanglement, hence the non-trivial time dependence.   Results are shown for various coupling strengths as indicated in the legend; other parameters as in Fig.~\ref{Fig:1}.
    Best fit exponential decays are plotted as dashed lines, to measure the time constant of decay to the HMF Gibbs state. Deviations from exponential decay at late times are caused by numerical errors in the simulations.}\label{Fig:2}
\end{figure}

To investigate this slow exponential decay further, Fig.~\ref{Fig:2} shows a fit of the late time behavior to an exponential decay with time constant $t_0$. In this figure we use a different initial state. Our initial state remains factorized, but the system part is prepared in the HMF Gibbs state.  Although this corresponds to the steady state of the reduced system density matrix, we nonetheless see time evolution due to the establishment of correlations between the system and the reservoir.  i.e. in contrast to the initial state, the steady state of the system and reservoir does not factorize.
The timescales extracted this way are shown in Fig.~\ref{Fig:3}, showing a rapid growth of $t_0$ with system--reservoir coupling $\alpha$, and a divergence of decay time as $\theta\to 0$.  One may note that, in agreement with~\cite{trushechkin2021quantum}, we see the diverging timescale for population relaxation in the ultrastrong-coupling limit implies the limits of $\alpha \to \infty$ and $t \to \infty$ do not commute; explaining the discrepancy of the results of Ref.~\cite{Goyal2019,Orman2020} in comparison with those of other works.

\subsection{Comparison to strong-coupling polaron theory}

The slow population decay and its dependence on $\alpha, \theta$, noted above for intermediate coupling, are consistent with the expectations at ultrastrong coupling, where one may use the polaron master equation or F\"orster  theory~\cite{Forster1948,McCutcheon2010,trushechkin2021quantum,trushechkin2021open}, which for completeness we summarize briefly here.  One starts by making a unitary polaron transform,
\begin{displaymath}
U=\exp\left[X \sum_k \frac{g_k}{\nu_k}(b_k-b^\dagger_k)\right],
\end{displaymath} 
and then treating the part of the system Hamiltonian which does not commute with $X$ perturbatively. 
In the current case, this means the we treat the term
$H_1=\omega_q \sin(\theta) \tau^x/2$ as a perturbation. This approximation is thus valid only when $\omega_q \sin(\theta) \ll \alpha \omega_c$.

When the perturbative approximation is valid, following standard methods~\cite{McCutcheon2010}, one can derive the polaron master equation which describes rates for transitions between the pointer states:
\begin{equation} \label{eq4}
\frac{d\rho}{dt}=
\frac{\Gamma_-}{2}
\Lagr[\tau_{-}]
+
\frac{\Gamma_+}{2}
\Lagr[\tau_{+}]
-i[H_S + H_{\text{Lamb}},\rho],
\end{equation}
where $H_{\text{Lamb}}$ describes Lamb shifts, and $\Gamma_\pm$ are transition rates.  The Lamb shift term can be ignored in the following, as it is diagonal in the $\tau^z$ basis, and so does not change the time evolution of $\langle \tau^z\rangle$.  
The rates for the transitions are given by
\begin{align} \label{eq3}
\Gamma_\pm&=\left(\frac{\omega_{q}\sin \theta}{2}\right)^{2}
\int^{\infty}_{-\infty}\text{d}s 
e^{\mp is\omega_q \cos\theta-4 C(s)}
\\
C(s)&=\int^{\infty}_{0}\text{d}\nu 
 \frac{J(\nu)}{\nu^{2}}\left[2\coth\left(\frac{\beta \nu}{2}\right)\sin^2\left(\frac{\nu s}{2}\right)
+i\sin \nu s\right]\nonumber
\end{align}

These rates obey the Kennard-Stepanov relation~\cite{Kennard1918,Kennard1926,Stepanov1957}, that
$\Gamma_+=\Gamma_- e^{-\beta \Delta E}$ where $\Delta E$ is the energy difference between the pointer states, $\Delta E=\omega_q \cos\theta$.  
This relation can be proven from the observation that, as a thermal equilibrium correlation function $C(s)$ obeys the Kubo-Martin-Schwinger~\cite{Kubo1957,Martin1959} condition $C(s)=C(-s-i\beta)$.

\begin{figure}[ht]
    \includegraphics[width=\linewidth]{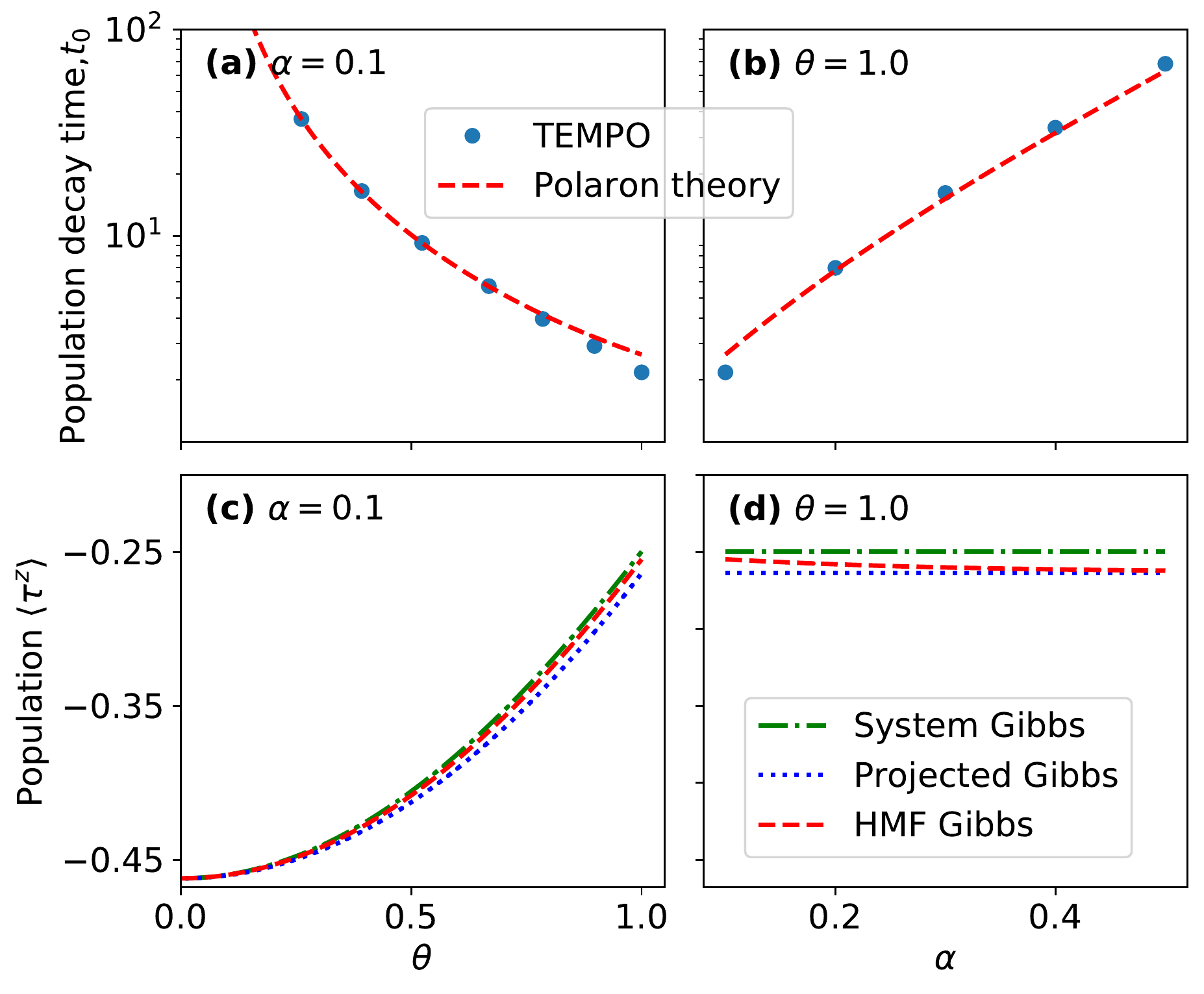}
    \caption{(a)-(b) Comparison of time constant $t_{0}$ found from TEMPO and polaron theory, as a function of (a) $\theta$ and (b) $\alpha$ as log-linear plots. (c)-(d) Steady state predictions of pointer-state population from HMF Gibbs state (as would be reached by TEMPO), system Gibbs state,  and projected system Gibbs state (as would be reached by polaron theory) in linear-linear plots.  
    Except where indicated, other parameters as in Fig.~\ref{Fig:1}.
    }\label{Fig:3}
\end{figure}

Because the polaron master equation describes transitions between pointer states $\ket{\tau=\uparrow,\downarrow}$, the steady state density matrix is diagonal in the pointer basis, and the probabilities $P_{\uparrow,\downarrow}$ obey $\partial_t P_{\uparrow}=- \partial_t P_{\downarrow}=-\Gamma_- P_{\uparrow} + \Gamma_+ P_{\downarrow}$, which thus gives the steady state $P_\uparrow/P_\downarrow = \Gamma_+/\Gamma_-=e^{-\beta \omega_q \cos\theta}$.  As such, this recovers the projected system state ensemble as its steady state~\cite{trushechkin2021quantum}.  One can also extract that the decay time of the populations towards this steady state is given by
\begin{equation} \label{eq5}
    \frac{1}{t_{0}}=  \Gamma_+ +\Gamma_-.
\end{equation}
Fig.~\ref{Fig:3}(a,b), compares the time constant $t_{0}$ found numerically to this prediction.  One sees that they match well when $\theta$ is small, and when $\alpha$ is large, consistent with the expectation of the condition $\omega_q \sin(\theta) \ll \alpha \omega_c$ for the polaron approach to be valid.

It should be noted that the polaron theory always predicts the projected system Gibbs state, so there exists a  difference between the true HMF Gibbs state and the polaron curve at intermediate coupling strengths.  This is shown in Fig.~\ref{Fig:3}(c,d). Both panels show that projected steady state (as predicted by the polaron master equation) differs from the HMF Gibbs state in general, but that the projected system Gibbs state converges to HMF Gibbs state in the limits where the polaron master equation is valid.

\section{Reservoir-induced coherence for two qubits}
\label{sec:TwoQubit}
In this section we demonstrate the usefulness of imaginary-time TEMPO by analyzing the HMF Gibbs state of a system of two qubits coupled to a common reservoir. We consider a system Hamiltonian $H_S=({\omega_q}/{2})(\sigma^z_a+\sigma^z_b)$ where the subscripts $a,b$ label the two qubits. The reservoir and interaction Hamiltonian is as in Eq.~\eqref{eq:coup}, but now with $X=(\sigma^x_a+\sigma^x_b)$.  We use the same Ohmic spectral density $J(\nu)$ as before.

This model has the feature that the coupling to the reservoir is expected to induce correlations between the qubits, but only at intermediate coupling strengths.  At weak coupling, the HMF Gibbs state approaches the system Gibbs state, with no correlation between the qubits.  At ultrastrong coupling, the HMF Gibbs state becomes the projected system Gibbs state;  in the current case, $H_{\text{proj}}=0$, so this yields a completely mixed state $\rho_{\text{proj}}=\mathbbm{1}/4$.

\begin{figure}[ht]
    \includegraphics[width=\linewidth]{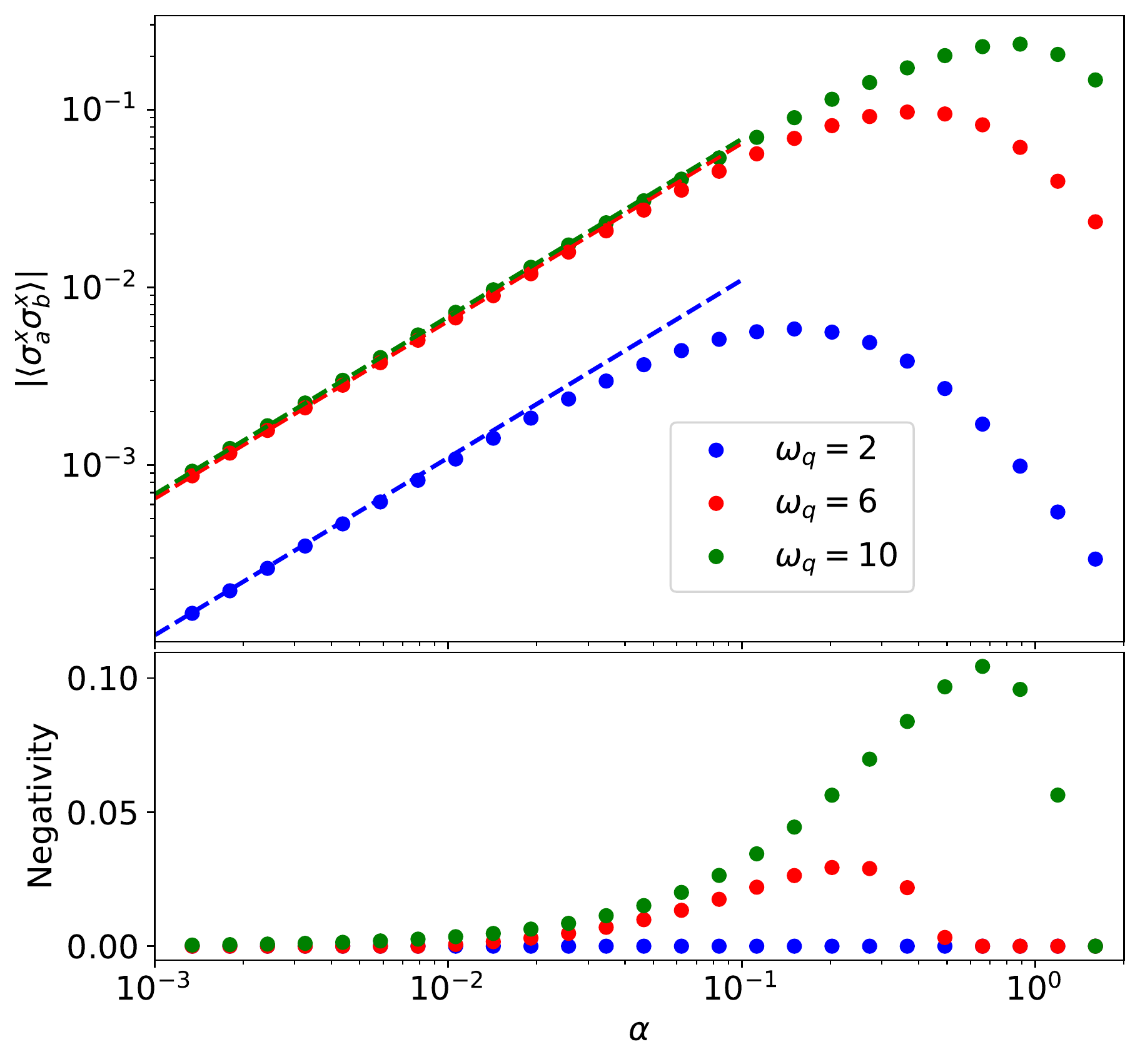}
    \caption{Qubit correlations vs system--reservoir coupling strength $\alpha$ in log-log plots.
    The top panel shows the absolute value of the $\langle \sigma^x_a \sigma^x_b\rangle$ coherence, for three different values of $\omega_{q}$. Dashed lines show the weak coupling analytical result of \citet{Cresser2021}. The bottom panel shows the negativity for each value of $\omega_{q}$. Other parameters are $k_{B}T=1, \omega_{c}=10$.}\label{Fig:4}
\end{figure}

In Figure~\ref{Fig:4} we show the evolution of correlations with system--reservoir coupling strength, measured in two different ways. First we consider the coherence $\expval{\sigma_{a}^{x}\sigma_{b}^{x}}$, noting that $\expval{\sigma^x_a}=\expval{\sigma^x_b}=0$. Secondly, to probe reservoir-induced entanglement in this mixed system we plot the negativity~\cite{Amico2008}. This is defined as the modulus of the sum of the negative eigenvalues of $\rho^{T_a}_{\text{HMF}}$, which is the HMF density matrix after partial transpose for qubit $a$.
As expected, we see correlations are strongest at intermediate coupling.
We may also note that the magnitude of the correlations, and the question of whether there is negativity, depends on the energy $\omega_q$.  Specifically, correlations are strongest when the system frequencies match the peak of the reservoir spectral density, i.e. when $\omega_q \simeq \omega_c$.  Indeed, when $\omega_q=2 \ll \omega_c$, while correlations are seen, the negativity remains zero for all couplings.

We also show the coherence from the weak-coupling analytical form, given by Eq.~(3) of Ref.~\cite{Cresser2021}. We see this agrees extremely well for $\alpha \leq 10^{-2}$. In the ultrastrong-coupling limit, as noted above, we expect $\rho_{\text{HMF}} \to \mathbbm{1}/4$.  As such, this is consistent with the correlations and negativity vanishing at large $\alpha$, but does not allow us to plot the limiting form.

\section{Conclusions}
\label{sec:Conclusions}

By using the Time Evolving Matrix Product Operator formalism for both real-time and imaginary time evolution, we have numerically confirmed the expected convergence of the open-system dynamics to the HMF Gibbs state. 
This shows that, as should be anticipated from general thermodynamic principles, the steady state of a system with intermediate or strong coupling to a reservoir is given by the HMF Gibbs state.  However, as noted elsewhere~\cite{trushechkin2021quantum}, the timescale for relaxation to this state diverges at ultrastrong coupling, implying a non-commuting limits of $t \to \infty$ and $\alpha \to \infty$.

We have also introduced imaginary-time TEMPO as a practical and efficient method to find the HMF Gibbs state numerically for intermediate coupling strength.  As this evolution acts directly in the Hilbert space, rather than a double Hilbert space, it is significantly more efficient than real-time propagation. We illustrate the potential of this method by considering reservoir-induced coherence between qubits.

\acknowledgments
We acknowledge helpful discussions with Janet Anders and Gerald Fux, and comments on a previous version of this manuscript from Janet Anders.
Y.F.C. acknowledges funding from the  St Andrews Undergraduate Research Assistant Scheme,  the School of Physics and Astronomy Student-Staff Council vacation awards, and the University of St Andrews Physics Trust. 
J.K. acknowledges funding from EPSRC (EP/T014032/1). 

\appendix

\section{Convergence of real- and imaginary-time TEMPO}

\begin{figure}[htpb]
    \includegraphics[width=\linewidth]{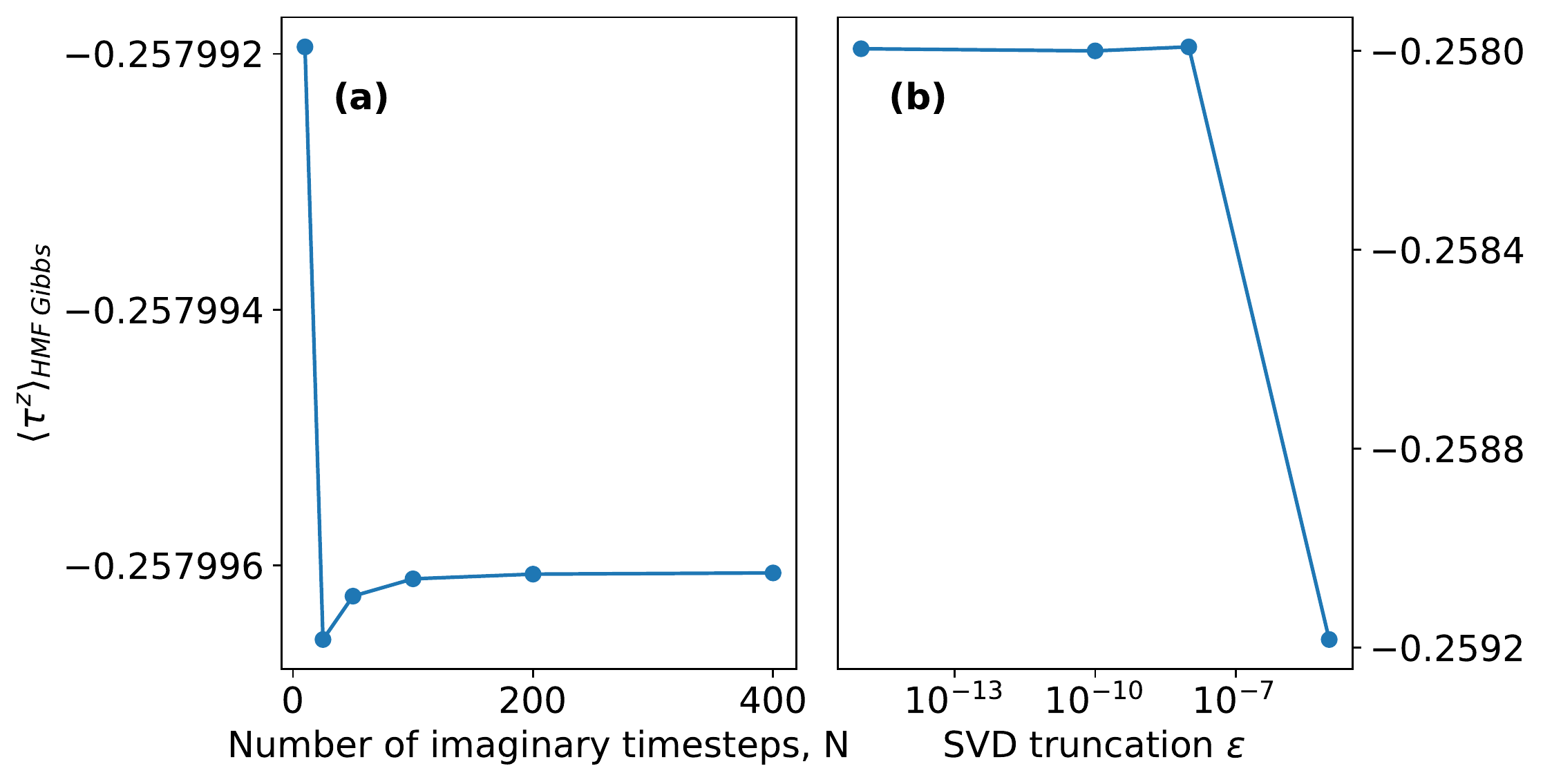}
    \caption{Convergence of the imaginary-time TEMPO calculation of the HMF Gibbs state with time step $d\tau$, as determined by $d\tau = \beta/N$.
    Plotted for $\alpha=0.2$, other parameters as in Fig.~\ref{Fig:1}. (a) shows convergence with respect to number of time step $N$ at a precision of $\epsilon=10^{-15}$ in linear-linear plot, and (b) shows convergence with respect to truncation precision $\epsilon$ at $N = 200$ in linear-log plot.
    }\label{AppenFig:2}
\end{figure}

In this section we present results demonstrating the numerical convergence of the real- and imaginary-time TEMPO algorithms, applied to the problem defined in Sec.~\ref{sec:OneQubit}.
The convergence of the HMF Gibbs state with respect to number of imaginary time steps $N$ and SVD truncation precision $\epsilon$ is shown in Fig.~\ref{AppenFig:2} . Noting the scale of the vertical axis, one sees that this shows very fast convergence with $N$ and $\epsilon$.  This convergence suggests high accuracy of the HMF Gibbs state used in the main text.

\begin{figure}[htpb]
    \includegraphics[width=\linewidth]{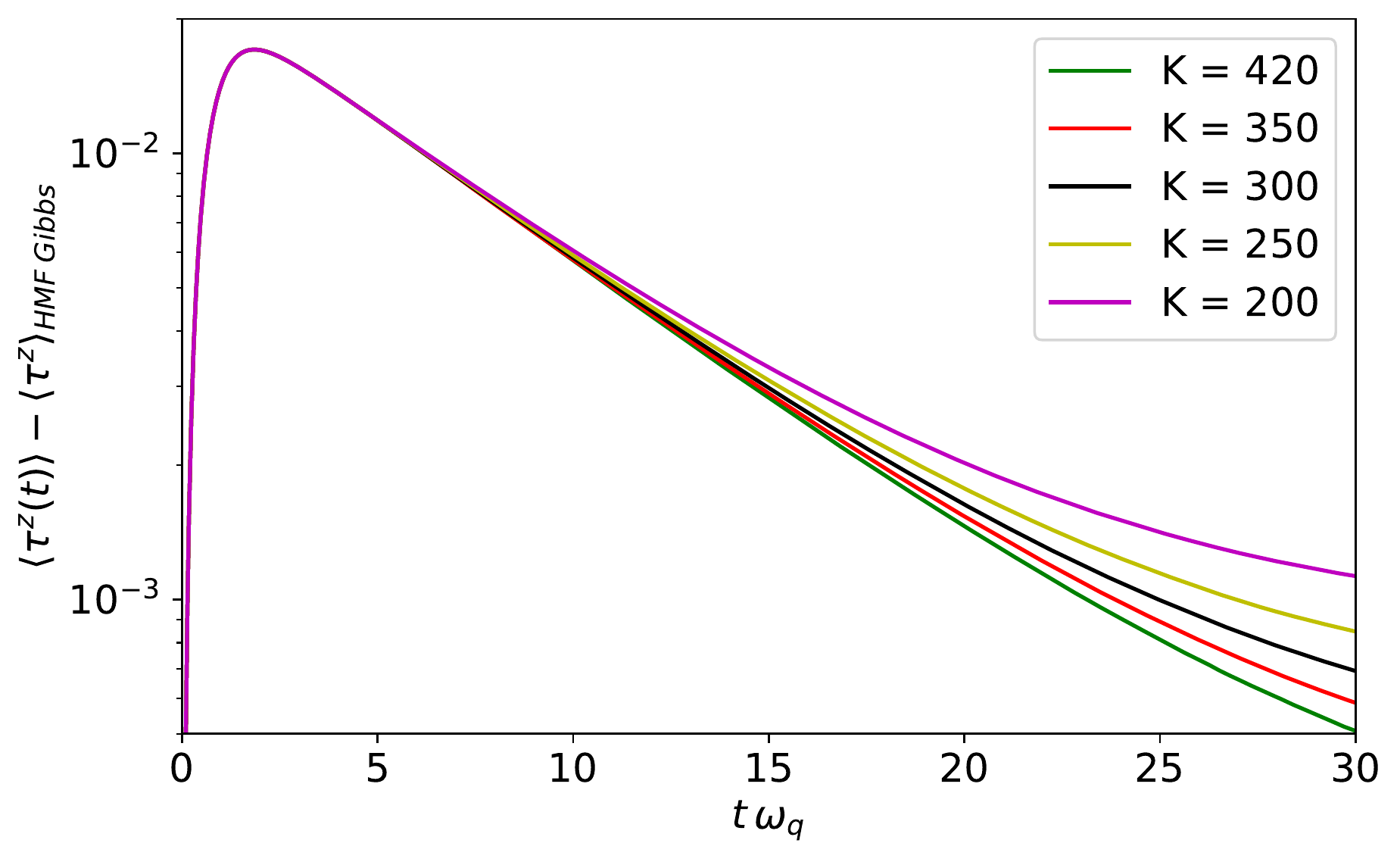}
    \caption{Convergence of real-time TEMPO with respect to memory length, $T_{\text{memory}}=K dt$ in log-linear plot. We show the difference between population in pointer basis and the HMF Gibbs state as in the main text.
    Plotted for $\alpha=0.2$, other parameters as in Fig.~\ref{Fig:1}.
    The simulation uses a time step $dt = 0.02$
    and SVD truncation of
    $\epsilon=10^{-9}$.}\label{AppenFig:1}
\end{figure}

In Fig.~\ref{AppenFig:1}, we show the dependence on the memory cutoff used in the real-time TEMPO algorithm.  To run to late times, we use a finite memory cutoff~\cite{Strathearn2018}, set by a parameter $K$, such that we neglect effects of the system more than $T_{\text{memory}}=K dt$ before the current time.  
We see convergence of the results of TEMPO as $K$ is increased.  We may note that the strongest effect of this memory cutoff is seen in the late time dynamics, as one approaches the steady state. 

\clearpage

\bibliography{tempo.bib}

\end{document}